\documentclass[twocolumn,trackchanges]{aastex63}

\usepackage{graphicx}
\graphicspath{{figures/}}
\usepackage{amsmath}
\usepackage{amssymb}
\usepackage{float}
\usepackage{color}
\usepackage{hyperref}

\usepackage{acro}

\DeclareAcronym{GR}{
	short = GR,
	long  = general relativity
	}
\DeclareAcronym{BH}{
	short = BH ,
	long  = black hole
}
\DeclareAcronym{BBH}{
	short = BBH ,
	long  = binary black hole
}
\DeclareAcronym{BNS}{
	short = BNS ,
	long  = binary neutron star
}
\DeclareAcronym{EFT}{
	short = EFT ,
	long  = effective field theory
}

\DeclareAcronym{GW}{
	short = GW ,
	long  = gravitational wave
}

\DeclareAcronym{CBC}{
	short = CBC,
	long  = compact binary coalescence
}

\DeclareAcronym{FRW}{
	short = FRW,
	long  = Friedmann-Robertson-Walker
}

\DeclareAcronym{PSD}{
	short = PSD,
	long  = power spectral density
}

\DeclareAcronym{LHC}{
	short = LHC,
	long  = Large Hadron Collider
}

\usepackage[capitalise]{cleveref}
\crefname{figure}{Fig.}{Figs.}
\Crefname{figure}{Fig.}{Figs.}

\def\be{\begin{equation}}
\def\ee{\end{equation}}
\def\({\left(}
\def\){\right)}
\def\[{\left[}
\def\]{\right]}

 \newcommand{\bqn}{\begin{eqnarray}}
 \newcommand{\eqn}{\end{eqnarray}}

 \newcommand{\f}{\frac}
 \newcommand{\p}{\partial}

\begin{document}
\title{Gravitational-Wave Implications for the Parity Symmetry of Gravity in the High Energy Region}

\author[0000-0002-2928-2916]{Yi-Fan Wang}
\email{yifan.wang@aei.mpg.de}
\affiliation{Max-Planck-Institut f{\"u}r Gravitationsphysik (Albert-Einstein-Institut), D-30167 Hannover, Germany}
\affiliation{Leibniz Universit{\"a}t Hannover, D-30167 Hannover, Germany}

\author{Rui Niu}
\affiliation{CAS Key Laboratory for Research in Galaxies and Cosmology, Department of Astronomy, University of Science and Technology of China, Hefei 230026, China}
\affiliation{School of Astronomy and Space Sciences, University of Science and Technology of China, Hefei, 230026, China}

\author{Tao Zhu}
\affiliation{Institute for theoretical physics and Cosmology, Zhejiang University of Technology, Hangzhou, 310032, China}
\affiliation{United center for gravitational wave physics (UCGWP), Zhejiang University of Technology, Hangzhou, 310032, China}

\author{Wen Zhao}
\email{wzhao7@ustc.edu.cn}
\affiliation{CAS Key Laboratory for Research in Galaxies and Cosmology, Department of Astronomy, University of Science and Technology of China, Hefei 230026, China}
\affiliation{School of Astronomy and Space Sciences, University of Science and Technology of China, Hefei, 230026, China}

\begin{abstract}
Einstein's general relativity, as the most successful theory of gravity, is one of the cornerstones of modern physics. 
However, the experimental tests for gravity in the high energy region are limited.
The emerging gravitational-wave astronomy has opened an avenue for probing the fundamental properties of gravity in strong and dynamical field, and in particular, high energy regime.
In this work, we test the parity conservation of gravity with gravitational waves. 
If the parity symmetry is broken, the left- and right-handed modes of gravitational waves would follow different equations of motion, dubbed as birefringence.
We perform full Bayesian inference by comparing the state-of-the-art waveform with parity violation with the compact binary coalescence data released by LIGO and Virgo collaboration.
We do not find any violations of general relativity, thus constrain the lower bound of the parity-violating energy scale to be $0.09$ GeV through the velocity birefringence of gravitational waves.  
This provides the most stringent experimental test of gravitational parity symmetry up to date.
We also find third generation gravitational-wave detectors can enhance this bound to $\mathcal{O}(10^2)$ GeV if there is still no violation, comparable to the current \ac{LHC} energy scale in particle physics, which indicates gravitational-wave astronomy can usher in a new era of testing the ultraviolet behavior of gravity in the high energy region.

\end{abstract}

\section{Introduction}
Symmetry is an essential characteristic of the fundamental theories of modern physics and thus necessary to be tested experimentally.
In this work, we focus on the parity symmetry, which indicates the invariance of physical laws under reversed spatial coordinates.
It is well-known that parity is conserved for strong and electromagnetic interactions but is broken in the weak interaction as firstly confirmed by the beta-decay experiment in cobalt-60 \citep{lee-yang,cswuparity}. 
Gravitational parity is conserved in Einstein's \ac{GR}. 
Nevertheless, various parity-violating gravity models, including Chern-Simons gravity \citep{cs}, ghost-free scalar-tensor gravity \citep{ghost}, the symmetric teleparallel equivalence of GR theory \citep{tele} and Ho\v{r}ava-Lifshitz gravity \citep{HL0,HL} have been proposed to account for the nature of dark energy, dark matter, or quantizing gravity. 
In particular, in some fundamental theories of gravity, such as string theory and loop quantum gravity, the parity violation in the high energy regime is inevitable \citep{cs}. 
Also note that, violation of gravity parity would indicate violation of Lorentz and CPT symmetry, which are expected to be conserved for all fundamental theories \citep{Kostelecky:2003fs,Kostelecky:2008ts,Kostelecky:2017zob}.

Various astrophysical tests have put lower limits on the parity violation energy scale.
Constraints from solar system test and binary pulsar observation are given by \cite{solar} and \cite{pulsar,pulsar2}, respectively, in the context of Chern-Simons gravity.
However, the observational evidence for gravity in the \textit{high energy} scale is limited, which leaves \ac{GW} observation as a last resort \citep{miller_2019,a1,a2}. 
In contrast to the tests of the solar system or binary pulsars, \ac{GW} reflects the wave behavior of the gravitational field.
The tiny deviation from \ac{GR}, if it exists, could be accumulated and magnified during propagation of \acp{GW}. 
For testing parity of gravity, \cite{mewes} inspected the peak split of \ac{GW} waveform due to birefringence in general Lorentz and parity-violating gravity , and put constraints with the \ac{GW} event GW150914 \citep{GW150914}.
\cite{1809} derived the \ac{GW} speed for generic parity-violating gravity and constrained the violation by the \ac{GW} speed measurement from GW180817/GRB170817A \citep{gw170817-speed}.
A quantitative summary is given in \cref{fig:multi_mpv} in section \ref{ch:result}.

Our work advances in the following aspects. 
We first derive the \ac{GW} waveform for generic parity-violating gravity from \ac{CBC} using our recent results \citep{zhao2020}.
Then the waveform is match-filtered with realistic \ac{GW} data within Bayesian inference framework.
This match-filtering process represents the maximum information we can extract from an individual \ac{GW} event to constrain parity violation.
In addition, the Bayesian approach allows us to combine constraints from multiple \ac{GW} events for a tighter result.
We obtain the most stringent constraints up to date for gravitational parity conservation at no higher than $0.09$ GeV, which demonstrates the feasibility of probing the \textit{high energy} behavior of gravity through \acp{GW}.

In what follows, we present our methods and results for inferring the constraints on parity violation in gravity from \ac{GW} measurements. 
We first introduce the construction of the parity-violating \ac{GW} waveform in the \ac{EFT} framework, then discuss the Bayesian inference for obtaining the constraints and the results. 
We also forecast the constraining ability of future ground-based \ac{GW} detectors.
It shows the third generation detectors such as Einstein Telescope and Cosmic Explorer can probe the parity violation energy scale to be $\mathcal{O}(10^2)$ GeV, comparable to the Large Hadron Collider (LHC) energy scale in particle physics.

\section{Waveform of gravitational waves with parity violation}
We first construct the generalized \ac{GW} waveform generated by \ac{CBC} with parity violation within the \ac{EFT} formalism.
\ac{EFT} provides a systematic framework to encode all kinds of modifications to an existing theory that could arise given certain new physics, thus simultaneously testing a range of modified gravity theories at once. 
To investigate the possible propagation effect due to parity violation, we consider the perturbation theory of gravitational field.
\ac{EFT} suggests that the leading-order modification to the  linearized action of \ac{GR} comes from two terms with three derivatives \citep{eft}, i.e., $\epsilon^{ijk}\dot{h}_{il}\partial_{j}\dot{h}_{kl}$ and  $\epsilon^{ijk}\partial^2{h}_{il}\partial_{j}{h}_{kl}$ with $\epsilon^{ijk}$ the antisymmetric symbol and $h_{ij}$ the tensor perturbation of metric, $\partial_j$ and a {\emph{dot}} denote the derivatives with respect to spatial coordinates and time, respectively, $\partial^2$ is the Laplacian, $i,j... = 1,2$ or $3$ refer to spatial coordinate.
Both terms are parity-violating.
Dimensional analysis dictates that these new terms are each suppressed by an energy scale.
We expect the two energy scales are of the same order, and denote collectively by $M_{\rm PV}$, which is the prime quantity we aim to constrain.
Otherwise, if the two energy scales differ by orders of magnitude, only the term with lower energy scale dominates, thus we can neglect another term and our result for $M_{\rm PV}$ will not change.

Thus, in the \ac{FRW} universe, choosing the unitary gauge, the linearized quadratic action of \ac{GR} with leading-order parity violation is \citep{eft}
\be\label{eq:action}
\begin{split}
S = \frac{1}{16 \pi G}\int dtd^3x a^3\Bigg[\frac{1}{4}\dot{h}_{ij}^{2}- \frac{1}{4a^2}(\partial_k h_{ij})^2  + \\
\frac{1}{4}\( \frac{c_1}{aM_\mathrm{PV}}\epsilon^{ijk}\dot{h}_{il}\partial_j \dot{h}_{kl} + \frac{c_2}{a^3M_\mathrm{PV} }\epsilon^{ijk}\partial^2 h_{il} \partial_j h_{kl}\) \Bigg] ,
\end{split}
\ee
where the last two terms with three derivatives correspond to the contribution from parity violation.
$c_1$ and $c_2$ are dimensionless coefficients, which are functions of cosmic time in general. 
As demonstrated by \cite{yunes2018}, the modifications to \ac{GR}-based \ac{GW} waveform only arise from the propagation effect given the leading-order parity violation modification discussed above, because the generation effect occurs on a radiation-reaction timescale much smaller than the \ac{GW} time of flight and its impact on the evolution of the \ac{GW} waveform is negligible.
We do not consider any modifications to \ac{GR} quadratic terms in \cref{eq:action}, such as Horndeski-like terms \citep{Horndeski:1974wa,1809,zhao2020} which do not violate parity symmetry.

This \ac{EFT} with leading-order extensions to \ac{GR} can be mapped to all the existing specific parity-violating modified gravity models  in the market. 
A detailed mapping dictionary can be found in \cite{zhao2020}. 
Note that, in EFT, the leading-order modifications from \textit{parity-conserving} terms in action are suppressed by $M_{\rm PV}^{-2}$, which indicates a much looser constraint of $M_{\rm PV}$ from \ac{GW} observation \citep{eft,mewes,zhao2020}.
In other words, since parity violation emerges at the leading-order modification, we expect the most stringent test on gravity from the propagation of \ac{GW} is on the gravitational parity symmetry from the viewpoint of \ac{EFT}.

Given the parity-violating terms,  the equation of motion for the \ac{GW} circular polarization mode $h_A$ in an \ac{FRW} universe is
\be\label{eq:eom}
h_A'' +(2+\nu_A)\mathcal{H}h_A' + (1+\mu_A)k ^2 h_A = 0~,
\ee
where $A=L$ or $R$ stands for the left- and right-hand modes, $\mathcal{H}$ is the conformal Hubble parameter, $k$ is the wavenumber, a {\emph{prime}} denotes the derivative with respect to the conformal cosmic time. 
Note that $\mu_A = \nu_A = 0$ would reduce \cref{eq:eom} to \ac{GR}.
Dimensional analysis indicates that both terms relate to the energy scale $M_{\rm PV}$ by $\mu_A  \propto \rho_A {k}/{M_{\rm PV}}$ and $\nu_A \propto \rho_A {k}/{M_{\rm PV}}$ with $\rho_R= 1$ and $\rho_L = -1$.
The broken parity leads to asymmetry between the left- and right-hand circular polarization modes of \ac{GW} during propagation.
In particular, the opposite sign of $\mu_A$ (as well as $\nu_A$) for different modes leads to the birefringence effect of \acp{GW}, which is a characterized phenomenon for \ac{GW} propagation in the parity-violating gravity. 
We find that the propagation of \acp{GW} can be affected in two ways. 
The term $\mu_A$ modifies the conventional dispersion relation of \acp{GW}. 
As a result, the velocities of left- and right-hand circular polarization of \acp{GW} are different, dubbed as the {\emph{velocity birefringence}} of \ac{GW} \citep{mewes}. 
On the other hand, the term $\nu_A$ induces the different damping rates for two polarization modes when they propagate in the expanding universe, which is called the {\emph{amplitude birefringence}} of \acp{GW} \citep{yunes2018}. 
In the general parity-violating gravity theories, both effects exist.
For each circular polarization mode, the former effect exactly induces the phase modifications of the \ac{GW} waveform, and the latter induces the amplitude modifications. 
Constraints on modification of equation of motion of the same type of \cref{eq:eom} are also obtained by LIGO and Virgo collaboration in \cite{GWTC1-testingGR}, but only parity-conserving terms are considered.
In contrast, our work focus on the parity-violating effect. 

The exact forms for $\mu_A$ and $\nu_A$ are \citep{zhao2020}
\bqn\label{coes2}
\nu_{A}&=& [\rho_{A}\alpha_{\nu}(\tau)(k/a M_{\rm PV})]'/\mathcal{H}, \nonumber \\ 
\mu_{A}&=&\rho_{A}\alpha_{\mu}(\tau)(k/a\ {M_{\rm PV}}),
\eqn
where $\tau$ is the cosmic conformal time. 
The functions $\alpha_{\nu}\equiv -c_1$ and $\alpha_{\mu}\equiv c_1-c_2$ are two arbitrary functions of time which can only be determined given a specific model of modified gravity.
For \ac{GW} events at the local universe, these two functions can be approximately treated as constant, i.e. ignoring their time-dependence.
We also let $\alpha_\mu$ and $\alpha_\nu$ to be $\sim\mathcal{O}(1)$ by absorbing the order of magnitude into $M_{\rm PV}$. 

The explicit parity-violating \ac{GW} waveform can be derived from solving the equation of motion.
Converting the left- and right-hand \ac{GW} polarization modes into the {\emph{plus}} and {\emph{cross}} modes which are used more often in \ac{GW} data analysis, the parity-violating waveform is
\bqn \label{eq:pvwaveform}
h_+^{\rm PV}(f) &=& h_+^{\rm GR}(f) - h_\times^{\rm GR}(f) (i\delta h - \delta \Psi), \nonumber \\
h_\times^{\rm PV}(f) &=& h_\times^{\rm GR}(f) + h_+^{\rm GR}(f)(i\delta h - \delta \Psi).
\eqn
The amplitude and phase modifications to the \ac{GR}-based waveform $h^{\rm GR}(f)$ due to birefringence take the following parametrized form
\bqn\label{eq:deltahpsi}
    \delta h(f)= 
    -A_\nu  \pi f  , ~~
   \delta \Psi(f)= 
   A_\mu (\pi f)^2/H_0,
\eqn
where $H_0$ is the Hubble constant. 
The coefficients
$A_\nu$ and $A_\mu$ are given by 
\bqn\label{eq:mpvconversion}
    A_\nu&=& 
    M_{\rm PV}^{-1} \( \alpha_\nu(z=0) - \alpha_\nu(z)(1+z)\), \nonumber \\
   A_\mu&=&
   M_{\rm PV}^{-1} \int^{z}_{0} \frac{\alpha_\mu(z') (1+z')dz'}{\sqrt{\Omega_M(1+z')^3 + \Omega_\Lambda}},
\eqn
where $z$ is the cosmic redshift.
We adopt a Planck cosmology ($\Omega_{M}=0.315$, $\Omega_{\Lambda}=0.685$, $H_0=67.4~{\rm km~s^{-1} ~Mpc^{-1}}$) \citep{Planck2018}. 

The above Eq.~(\ref{eq:pvwaveform}) represents the waveform we employ to compare with the \ac{GW} data. 
Assume the \ac{GW} is emitted at the redshift $z\sim O(0.1)$, and $\alpha_{\nu}$ and $\alpha_{\mu}$ are expected to be the same order constant, we find the ratio $\delta\Psi/\delta h \sim \pi f/H_0\gtrsim 10^{20}$, where $f\sim 100$ Hz for the ground-based \ac{GW} detectors. 
Therefore, in the general parity-violating gravity, the corrections of \ac{GW} waveform ${h}^{\rm PV}(f)$ mainly come from the contribution of velocity birefringence rather than that of amplitude birefringence.
Note that, though we choose $\alpha_\nu=c_1-c_2 = 1$ by absorbing its order of magnitude into $M_\mathrm{PV}$, it is possible that $\alpha_\nu=0$ as is the case for a particular parity-violating gravity, the Chern-Simons gravity.
Nevertheless, $\alpha_\nu$ is non-zero for more general parity-violating gravity constructed from \ac{EFT} \citep{ghost,zhao2020} and corresponding to ghost-free scalar-tensor gravity \citep{ghost}, the symmetric teleparallel equivalence of GR theory \citep{tele} or Ho\v{r}ava-Lifshitz gravity \citep{HL0,HL}.
Our main results correspond to non-zero $\alpha_\nu$, but we also consider the constraint for Chern-Simons gravity as a separate case, which is equivalent to let $\delta\Psi=0$ in Eq.~(\ref{eq:pvwaveform}).


\section{Bayesian inference for gravitational-wave events}

With the waveform Eq.~(\ref{eq:pvwaveform}), we can perform a direct comparison with the \ac{GW} data using Bayesian inference to test the parity violation.
At the time of writing, LIGO and Virgo collaborations have released the data of confident \ac{CBC} events from the catalog GWTC-1 \citep{gwtc} which include ten \ac{BBH} events and a \ac{BNS} event GW170817, together with a second \ac{BNS} event, GW190425 \citep{GW190425}.
We analyze the open data \citep{gwosc} of the twelve events with the inference module of the open-source software \texttt{PyCBC}  \citep{pycbcinference} developed for \ac{GW} astronomy, which in turn has dependency on \texttt{LALSuite} \citep{lalsuite}.

Bayesian inference framework is broadly employed in \ac{GW} astronomy for estimating the source parameters and selecting the preferred model from observation.
Given the data $d$ of \ac{GW} signal and a waveform model $H$, Bayes theorem claims
\be\label{eq:pe}
P(\vec{\theta}|d,H,I) = \frac{ P(d|\vec{\theta},H,I) P(\vec{\theta}|H,I)} {P(d|H,I)},
\ee
where $\vec{\theta}$ are the parameters characterizing $H$, $I$ is any other background information. 
$P(\vec{\theta}|H,I)$ is the prior distribution for $\vec{\theta}$ and $P(d|\vec{\theta},H,I)$ is the likelihood for obtaining the data given a specific set of model parameters.
The posterior $P(\vec{\theta}|d,H,I)$ contains all the information about the results of parameter estimation. 
To combine the posterior from each event to give an overall inference is also straightforward
\be
p(\vec{\theta}|\{d_i\},H, I) \propto \prod_{i=1}^{N}  p(\vec{\theta}|d_i, H, I),
\ee
where $d_i$ denotes the $i$-th \ac{GW} event.

For Gaussian and stationary noise from \ac{GW} detectors, the likelihood function reads
\be \label{eq:likelihood}
P(d|\vec{\theta}, H, I) \propto \exp\[	-\frac{1}{2}\sum_{i} \langle d-h(\vec{\theta})|d-h(\vec{\theta})\rangle \],
\ee
where $h(\vec{\theta})$ is the \ac{GW} waveform template in model $H$, and $i$ represents the $i$-th \ac{GW} detector.
The inner product $\langle a|b\rangle$ is defined to be
\be
\langle a|b\rangle = 4 \mathfrak{R} \int \frac{{a}(f){b}^*(f)}{S_h(f)} df,
\ee
where $S_h(f)$ is the one-side noise \ac{PSD} of the \ac{GW} detector.

To select the model favored by observation, normalizing both sides of \cref{eq:pe} yields the Bayes evidence
\be 
P(d|H,I) = \int d\vec{\theta} P(d|\vec{\theta},H,I) P(\vec{\theta}|H,I).
\ee
Bayesian ratio is defined as the ratio of evidence between two competitive models $H_1$ and $H_2$ which are \ac{GR} and parity-violating gravity within this work,
\be 
\mathcal{B}^1_2 = \frac{P(d|H_1,I)}{P(d|H_2,I)}.
\ee
The odds ratio between model $H_1$ and model $H_2$ can be expressed by
\be \label{eq:oddsratio}
\mathcal{O}^1_2 = \frac{P(H_1|d,I)}{P(H_2|d,I)} = \frac{P(d|H_1,I)}{P(d|H_2,I)} \frac{P(H_1|I)}{P(H_2|I)} = \mathcal{B}^1_2 \frac{P(H_1|I)}{P(H_2|I)}.
\ee
Odds ratio is equal to Bayesian ratio if the competitive models are assumed to be equally likely before any measurement, which quantitatively reflects the preference of data for competitive models.

We employ the open-source software \texttt{PyCBC} with the open data from \cite{gwosc} to perform the  Bayesian inference.
For the \ac{GR} waveform $h^{\rm GR}(f)$, we use the spin precessing waveform \texttt{IMRPhenomPv2} \citep{pv2,pv22} when analyzing \ac{BBH} events and spin aligned waveform with tidal deformability \texttt{IMRPhenomD\_NRTidal} \citep{nrtidal} for \ac{BNS} events. 
The parity-violating waveform is constructed based on the above template through Eq.(\ref{eq:pvwaveform}).
We perform parameter estimation by selecting 16s data for \ac{BBH} and 200s data for \ac{BNS} events to account for the relatively long signal.
The data is sampled at 2048 Hz and the likelihood is evaluated between 20 and 1024 Hz.
The \ac{PSD} is generated from 1000s data using the median estimation with 8s Hann-windowed segments and overlapped by 4s \citep{pycbcinference}.
The prior is chosen to be consistent with that of \cite{gwtc} and uniformly distributed for $A_\mu$ and $A_\nu$.
The posterior distribution is sampled by the nest sampling algorithm $\texttt{dynesty}$ \citep{dynesty} over the fiducial \ac{BBH} and \ac{BNS} source parameters plus the parity-violating parameters $A_\mu$ for velocity birefringence or $A_\nu$ for amplitude birefringence.

\section{Results of Constraints on Parity Violation}\label{ch:result}
\begin{figure}
\includegraphics[width=\columnwidth]{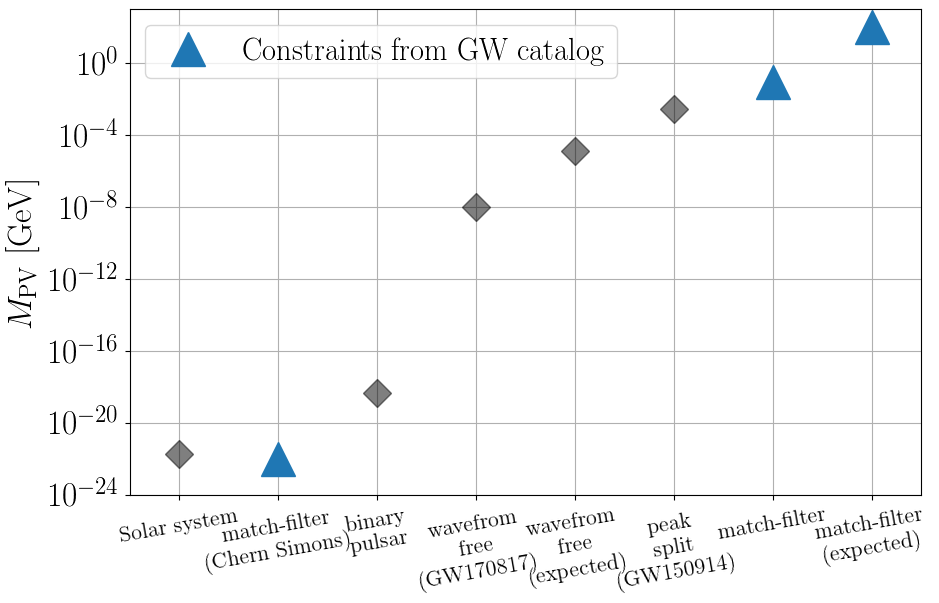}
\caption{{\bf Constraints on the lower limits of parity-violating energy scales in gravity from various observations.} 
The triangle markers denote the constraints by waveform match-filter with the \ac{GW} catalog for the general parity-violating gravity with amplitude and velocity birefringence and the special case with only amplitude birefringence (e.g., Chern Simons gravity), respectively.
The last column represents the expected constraints for general parity violation with match-filter method with the third generation ground-based \ac{GW} detectors.
Other existing constraints including the Solar system tests \citep{solar}, the binary pulsar observation \citep{pulsar,pulsar2}, the waveform-free result using \ac{GW} speed measurements from GW170817 \citep{1809}, future expected result with \ac{GW} speed measurements \citep{Zhao:2019}, inspection of peak split with GW150914 \citep{mewes} are also plotted for comparison. }
\label{fig:multi_mpv}
\end{figure}

For all the \ac{GW} events, we do not find any signatures of parity violation.
We thus put the lower limit of $M_{\rm PV}$ to be $0.09$ GeV in the general parity-violating gravity, which is the most stringent limit up to date.
The results on $M_{\rm PV}$ are shown in \cref{fig:multi_mpv}, where we have combined the results from the twelve CBC events to give an overall constraint.
For comparison, the figure also includes the results from the LAGEOS satellite in the solar system \citep{solar}, that from the double pulsar system PSR J0737-3039 A/B \citep{pulsar2}, that from the waveform-free method using \ac{GW} speed measurement from arrival time difference between GW170817 and GRB170817A \citep{1809}, the expected result by considering the waveform-free method developed in \cite{Zhao:2019}, and constraints from the inspection of no peak split with GW150914 in \cite{mewes}.

We note that this result has direct application on constraining a range of specific parity-violating gravity models with velocity birefringence, including the ghost-free scalar-tensor gravity \citep{ghost}, the symmetric teleparallel equivalence of GR theory \citep{tele} and Ho\v{r}ava-Lifshitz gravity \citep{HL0,HL}.
The detailed correspondence between the above modified gravity models and the \ac{EFT} formalism can be found in Ref.~\citep{zhao2020}.
The constraints can also be mapped to the standard-model extension \citep{Kostelecky:2008ts} framework for Lorentz and CPT-violated gravity, and the generalized framework for testing \ac{GR} with \ac{GW} propagation \citep{Nishizawa:2017nef,Arai:2017hxj,Nishizawa:2019rra}, but a detailed mapping relation is beyond the scope of this manuscript and thus left for future work.
We have made our inference results open \footnote{\url{https://github.com/yi-fan-wang/ParitywithGW}} to facilitate mapping the constraint to any specific parity-violating gravity theories that one is interested in.

By similar analysis, the constraint by only considering the amplitude birefringence modification is $M_{\rm PV}>1 \times10^{-22}$ GeV. 
This result can be directly compared to and is consistent with \cite{cs4,yunes2018,Nair:2019iur} which focus on testing Chern-Simons gravity with \ac{GW}.
Compared to the constraint for general parity violation, the loose result for amplitude birefringence is because $\delta h$ is negligibly small compared to $\delta \Psi$ and the \ac{GW} detection is less sensitive to amplitude modification than phase.

\begin{figure}[htbp]
\includegraphics[width=\columnwidth]{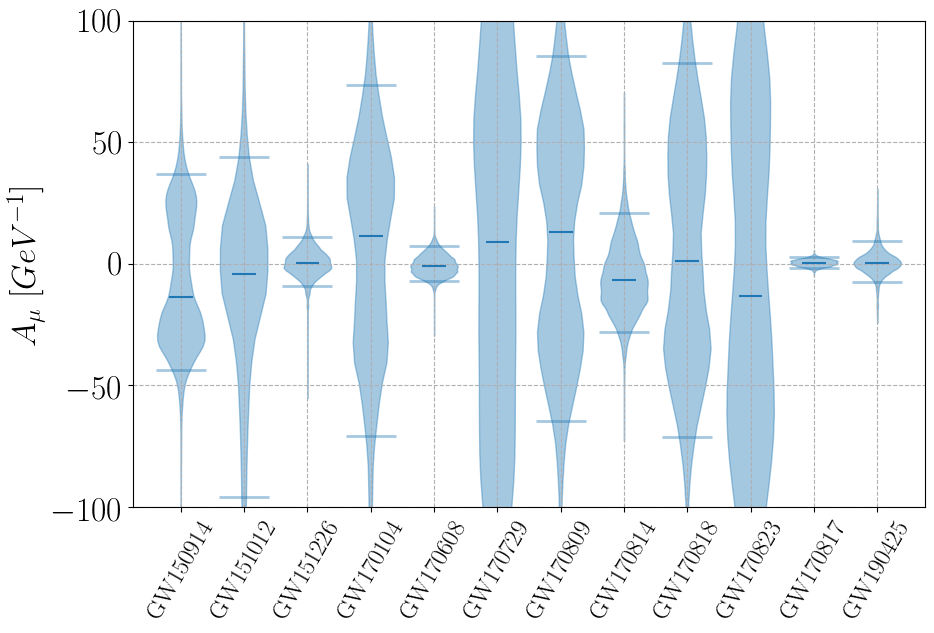}
\caption{{\bf Violin plots of the posteriors of the parameter $A_{\mu}$} 
The results are obtained by analyzing the twelve \ac{GW} events. The region in the posterior between the upper and lower bar denote the $90\%$ credible interval, and the bar at the middle denotes the median value.
The \ac{GR} value $A_\mu = 0$ is within the $90\%$ confidence interval for each event. 
We notice that the two relative low-mass events GW151226 and GW170608 and the two \ac{BNS} events yield the best constraint on $A_\mu$. }
\label{fig:violin}
\end{figure}

\begin{figure}[htbp]
\includegraphics[width=\columnwidth]{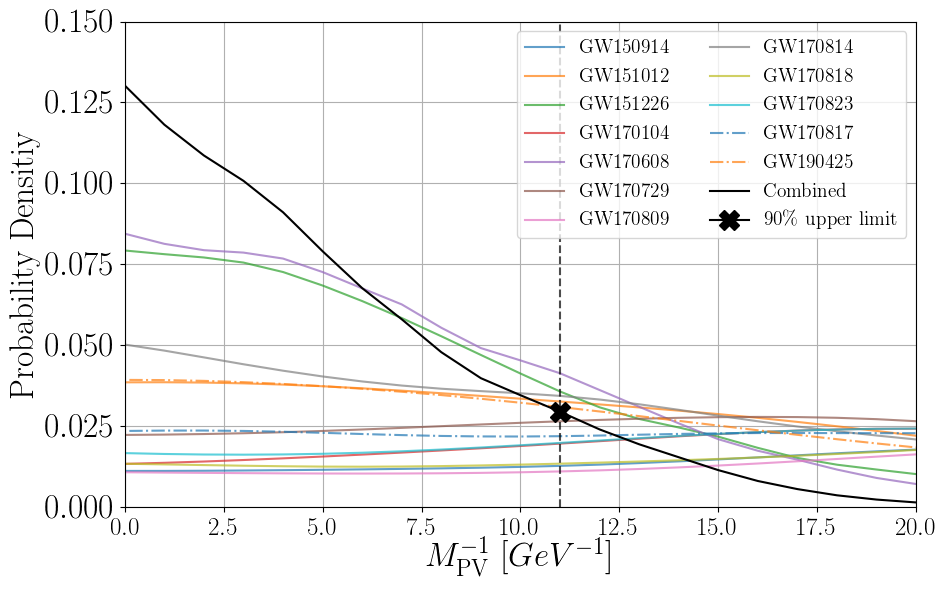}
\caption{{\bf The posterior distributions for $M_\mathrm{PV}^{-1}$}. 
The inference results for the parity-violating energy scale for velocity birefringence are plotted.
The results from an individual \ac{GW} event and the combination of all the events are considered.
The abscissa value of the ``x'' marker represents the $90\%$ upper limit for $M_\mathrm{PV}^{-1}$, or equivalently, the $90\%$ lower limit for $M_\mathrm{PV}$, which is $0.09$ GeV.}
\label{fig:single_event_mpv}
\end{figure}

For analyzing the results from each individual event, we also plot \cref{fig:violin} to show the marginalized posterior distribution for $A_\mu$ (velocity birefringence) and \cref{fig:single_event_mpv} for the marginalized posterior distributions of $M_{\rm PV}$.
From \cref{fig:violin}, we observe that the \ac{GR} value $A_\mu=0$ is within the $90\%$ confidence level for every event.
We also report that the natural logarithm of the Bayes ratio between \ac{GR} and the parity-violating gravity is in the range $[1.6, 5.8]$ for all the events, confirming no parity violation for gravity.

The relatively low-mass \ac{BBH} events, such as GW151226, GW170608, and the two \ac{BNS} events give tighter constraints on $A_\mu$. 
This is because the velocity birefringence contribution corresponds to a $5.5$ post-Newtonian (PN) order modification to the \ac{GR} waveform which has a larger impact on higher frequency, thus the low-mass events with higher cutoff frequency and longer signal yield better constraints.

In \cref{fig:single_event_mpv}, for converting $A_\mu$ into $M_{\rm PV}$, we absorb the absolute value of $\alpha_{\mu}$ into the definition of $M_{\rm PV}$, which is equivalent to setting $|\alpha_{\mu}|=1$ in the calculation, then the parameter $M_{\rm PV}$ can be obtained from $A_{\mu}$ and redshift $z$ by Eq.~(\ref{eq:mpvconversion}).
The combined results show that the $90\%$ lower limit for $M_{\rm PV}$ is $0.09$ GeV, representing the tightest constraint on $M_{\rm PV}$ up to date.

%
%

\section{Future constraints with more advanced detectors}

With the continuing sensitivity upgrade during the advanced LIGO and Virgo runs, the future \ac{GW} astronomy is even more powerful to test the parity symmetry of gravity.
The KAGRA detector has joined the global network very recently.
The advanced LIGO and Virgo detectors are expected to achieve the design sensitivity in a few years \citep{livingreviewligo}.
The third generation ground-based \ac{GW} detectors, including the Einstein Telescope and the Cosmic Explorer, are under projection currently \citep{ce}.
We investigate the ability of future \ac{GW} astronomy to constrain the lower limit of $M_{\rm PV}$ by simulations.

We consider four sets of detector configurations based on technologies currently available or under investigation, and simulate 200 \ac{BBH} events from \ac{GR} for each set.
For the first set, we choose the advanced LIGO, advanced Virgo and KAGRA network, all running with designed sensitivity.
The second and third sets substitute the two LIGO detectors with the 2.5 generation detector A+ \footnote{\url{https://dcc.ligo.org/LIGO-T1800042/public}} and the Voyager \footnote{\url{https://dcc.ligo.org/LIGO-T1500293-v11/public}} configuration, respectively.
The last set uses the third generation detectors including the Einstein Telescope \citep{et,ce} and two Cosmic Explorer \citep{cewhitepaper,ce} detectors located at the LIGO sites.

The simulated \ac{BBH} events are uniformly located in the space and have mass uniformly distributed in the range $[5,50] M_\odot$.
For the first three sets, the upper cutoff for luminosity distance is chosen to be $2000$ Mpc, while for the third generation detectors the distance cutoff is $5000$ Mpc.
We do not consider more distant sources to give a conservative estimation of the constraining ability of the future \ac{GW} detector configurations.
We employ Bayesian inference to perform parameter estimation on the simulated events and choose the signal-to-noise ratio $>8$ as the criterion for detection.

\begin{figure}[htbp]
\includegraphics[width=\columnwidth]{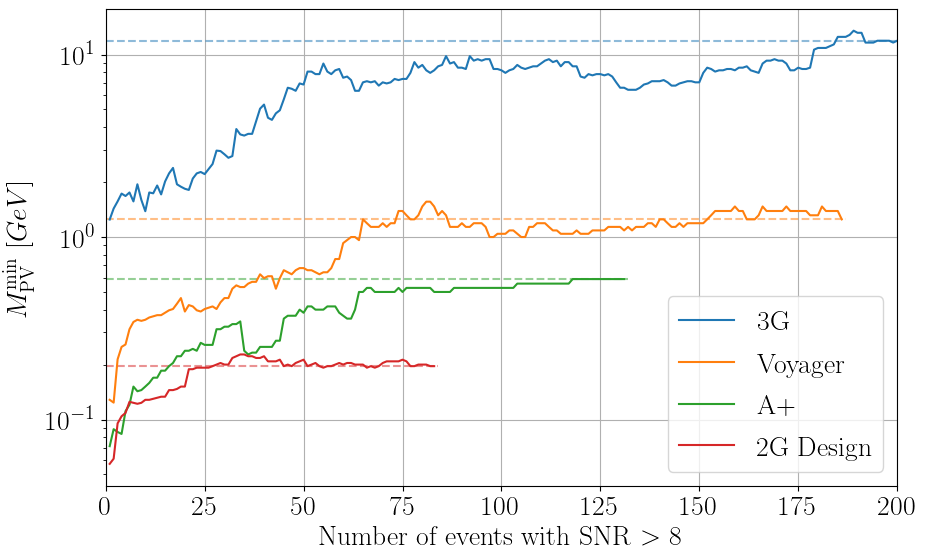}
\caption{{\bf The projected constraints for $M_{\rm PV}$ for future \ac{GW} detectors.} 
Using $200$ \ac{BBH} signals from \ac{GR}, the constraints for the lower limit of the parity-violating energy scale for velocity birefringence are plotted.
We consider four sets of global \ac{GW} detectors network, which are: (1) the second generation detectors including advanced LIGO, advanced Virgo, and KAGRA with design sensitivity; (2) the 2.5 generation detector A+; (3) the 2.5 generation detector Voyager; (4) the third generation detector with the Einstein Telescope and  Cosmic Explorer.
As the number of detections increase, the constraints for the lower limit of $M_{\rm PV}$ becomes tighter.
In particular, the third generation detector can detect all the \ac{BBH} coalescence signals within $5000$ Mpc and can constrain $M_{\rm PV} > \mathcal{O}(10)$ GeV with $200$ events.
With a one-year observation run, the third generation detectors are expected to improve the constraint to $\mathcal{O}(10^2)$ GeV.
}
\label{fig:3G}
\end{figure}

In \cref{fig:3G}, we show the results of the combined constraints on $M_{\rm PV}$ with respect to the number of detections. 
We first notice that, out of the 200 sources, $40\%$ of the sources can be detected by the advanced LIGO, advanced Virgo and KAGRA global network with design sensitivity, and $M_\mathrm{PV}$ can be constrained to $0.2$ GeV, while the 2.5 generation detectors A+ and Voyager can resolve $65\%$ and $90\%$ of sources, respectively, and constrain $M_\mathrm{PV}$ to be not less than $0.6$ GeV and $1$ GeV.

For the last set of simulations, all the \ac{BBH} sources can be detected by the third generation detectors, and the constraint with 200 events is $10$ GeV.
Given the local merger rate estimation $53.2$ yr$^{-1}$ Gpc$^{-3}$ from LIGO and Virgo \citep{GWTC1-rate}, it is expected that there are $\mathcal{O}({10^4})$ \ac{BBH} coalescence events within 5000 Mpc in one year.
Therefore, assuming the constraint on $M_\mathrm{PV}$ is inversely proportional to the square root of event number, the resultant constraint can reach $\mathcal{O}(10^2)$ GeV with a one-year observation with the third generation detectors.
This demonstrates the promising future of \ac{GW} astronomy to probe the ultraviolet property of gravity in the high energy region, which could shed light on deviations from \ac{GR}, if existing, arising from the $100$ GeV  region.

%
%

\acknowledgments
Y.-F.W. thanks Badri Krishnan, Collin Capano and Alex Nitz for the fruitful discussion and acknowledges the Max Planck Gesellschaft for support and the Atlas cluster computing team at AEI Hannover.
W.Z. and R.N. appreciate the helpful discussion with Linqing Wen, Xing Zhang, Qian Hu, Mingzheng Li and Anzhong Wang.
W.Z. is supported by NSFC Grants No. 11773028, No. 11633001, No. 11653002, No. 11421303, No. 11903030, and the Strategic Priority Research Program of the Chinese Academy of Sciences Grant No. XDB23010200. 
T.Z. is supported in part by NSFC grants No. 11675143, the Zhejiang Provincial NSFC grants No. LR21A050001 and No. LY20A050002, and the Fundamental Research Funds for the Provincial Universities of Zhejiang in China grants No. RF- A2019015.
This research has made use of data, software and/or web tools obtained from the Gravitational Wave Open Science Center (https://www.gw-openscience.org), a service of LIGO Laboratory, the LIGO Scientific Collaboration and the Virgo Collaboration. LIGO is funded by the U.S. National Science Foundation. Virgo is funded by the French Centre National de Recherche Scientifique (CNRS), the Italian Istituto Nazionale della Fisica Nucleare (INFN) and the Dutch Nikhef, with contributions by Polish and Hungarian institutes.

\vspace{5mm}

\bibliographystyle{aasjournal}
\bibliography{parity.bib}

\begin{thebibliography}{}
\expandafter\ifx\csname natexlab\endcsname\relax\def\natexlab#1{#1}\fi
\providecommand{\url}[1]{\href{#1}{#1}}
\providecommand{\dodoi}[1]{doi:~\href{http://doi.org/#1}{\nolinkurl{#1}}}
\providecommand{\doeprint}[1]{\href{http://ascl.net/#1}{\nolinkurl{http://ascl.net/#1}}}
\providecommand{\doarXiv}[1]{\href{https://arxiv.org/abs/#1}{\nolinkurl{https://arxiv.org/abs/#1}}}

\bibitem[{Abbott {et~al.}(2016)}]{GW150914}
Abbott, B.~P., {et~al.} 2016, Phys. Rev. Lett., 116, 061102,
  \dodoi{10.1103/PhysRevLett.116.061102}

\bibitem[{Abbott {et~al.}(2017{\natexlab{a}})Abbott, Abbott, Abbott, Acernese,
  Ackley, Adams, Adams, Addesso, Adhikari, Adya, Affeldt, Afrough, Agarwal,
  Agathos, Agatsuma, Aggarwal, Aguiar, Aiello, Ain, Ajith, Allen, Allen,
  Allocca, Aloy, Altin, Amato, Ananyeva, Anderson, Anderson, Angelova, Antier,
  Appert, Arai, Araya, Areeda, Arnaud, Arun, Ascenzi, Ashton, Ast, Aston,
  Astone, Atallah, Aufmuth, Aulbert, AultONeal, Austin, Avila-Alvarez, Babak,
  Bacon, Bader, Bae, Baker, Baldaccini, Ballardin, Ballmer, Banagiri, Barayoga,
  Barclay, Barish, Barker, Barkett, Barone, Barr, Barsotti, Barsuglia, Barta,
  Bartlett, Bartos, Bassiri, Basti, Batch, Bawaj, Bayley, Bazzan, B{\'{e}}csy,
  Beer, Bejger, Belahcene, Bell, Berger, Bergmann, Bero, Berry, Bersanetti,
  Bertolini, Betzwieser, Bhagwat, Bhandare, Bilenko, Billingsley, Billman,
  Birch, Birney, Birnholtz, Biscans, Biscoveanu, Bisht, Bitossi, Biwer,
  Bizouard, Blackburn, Blackman, Blair, Blair, Blair, Bloemen, Bock, Bode,
  Boer, Bogaert, Bohe, Bondu, Bonilla, Bonnand, Boom, Bork, Boschi, Bose,
  Bossie, Bouffanais, Bozzi, Bradaschia, Brady, Branchesi, Brau, Briant,
  Brillet, Brinkmann, Brisson, Brockill, Broida, Brooks, Brown, Brown, Brunett,
  Buchanan, Buikema, Bulik, Bulten, Buonanno, Buskulic, Buy, Byer, Cabero,
  Cadonati, Cagnoli, Cahillane, Bustillo, Callister, Calloni, Camp, Canepa,
  Canizares, Cannon, Cao, Cao, Capano, Capocasa, Carbognani, Caride, Carney,
  Diaz, Casentini, Caudill, Cavagli{\`{a}}, Cavalier, Cavalieri, Cella, Cepeda,
  Cerd{\'{a}}-Dur{\'{a}}n, Cerretani, Cesarini, Chamberlin, Chan, Chao,
  Charlton, Chase, Chassande-Mottin, Chatterjee, Chatziioannou, Cheeseboro,
  Chen, Chen, Chen, Cheng, Chia, Chincarini, Chiummo, Chmiel, Cho, Cho, Chow,
  Christensen, Chu, Chua, Chua, Chung, Chung, Ciani, Ciolfi, Cirelli, Cirone,
  Clara, Clark, Clearwater, Cleva, Cocchieri, Coccia, Cohadon, Cohen, Colla,
  Collette, Cominsky, Jr., Conti, Cooper, Corban, Corbitt,
  Cordero-Carri{\'{o}}n, Corley, Cornish, Corsi, Cortese, Costa, Coughlin,
  Coughlin, Coulon, Countryman, Couvares, Covas, Cowan, Coward, Cowart, Coyne,
  Coyne, Creighton, Creighton, Cripe, Crowder, Cullen, Cumming, Cunningham,
  Cuoco, Canton, D{\'{a}}lya, Danilishin, D'Antonio, Danzmann, Dasgupta, Costa,
  Dattilo, Dave, Davier, Davis, Daw, Day, De, DeBra, Degallaix, Laurentis,
  Del{\'{e}}glise, Pozzo, Demos, Denker, Dent, Pietri, Dergachev, Rosa, DeRosa,
  Rossi, DeSalvo, de~Varona, Devenson, Dhurandhar, D{\'{\i}}az, Fiore,
  Giovanni, Girolamo, Lieto, Pace, Palma, Renzo, Doctor, Dolique, Donovan,
  Dooley, Doravari, Dorrington, Douglas, {\'{A}}lvarez, Downes, Drago,
  Dreissigacker, Driggers, Du, Ducrot, Dupej, Dwyer, Edo, Edwards, Effler,
  Eggenstein, Ehrens, Eichholz, Eikenberry, Eisenstein, Essick, Estevez,
  Etienne, Etzel, Evans, Evans, Factourovich, Fafone, Fair, Fairhurst, Fan,
  Farinon, Farr, Farr, Fauchon-Jones, Favata, Fays, Fee, Fehrmann, Feicht,
  Fejer, Fernandez-Galiana, Ferrante, Ferreira, Ferrini, Fidecaro, Finstad,
  Fiori, Fiorucci, Fishbach, Fisher, Fitz-Axen, Flaminio, Fletcher, Fong, Font,
  Forsyth, Forsyth, Fournier, Frasca, Frasconi, Frei, Freise, Frey, Frey,
  Fries, Fritschel, Frolov, Fulda, Fyffe, Gabbard, Gadre, Gaebel, Gair,
  Gammaitoni, Ganija, Gaonkar, Garcia-Quiros, Garufi, Gateley, Gaudio, Gaur,
  Gayathri, Gehrels, Gemme, Genin, Gennai, George, George, Gergely, Germain,
  Ghonge, Ghosh, Ghosh, Ghosh, Giaime, Giardina, Giazotto, Gill, Glover, Goetz,
  Goetz, Gomes, Goncharov, Gonz{\'{a}}lez, Castro, Gopakumar, Gorodetsky,
  Gossan, Gosselin, Gouaty, Grado, Graef, Granata, Grant, Gras, Gray, Greco,
  Green, Gretarsson, Groot, Grote, Grunewald, Gruning, Guidi, Guo, Gupta,
  Gupta, Gushwa, Gustafson, Gustafson, Halim, Hall, Hall, Hamilton, Hammond,
  Haney, Hanke, Hanks, Hanna, Hannam, Hannuksela, Hanson, Hardwick, Harms,
  Harry, Harry, Hart, Haster, Haughian, Healy, Heidmann, Heintze, Heitmann,
  Hello, Hemming, Hendry, Heng, Hennig, Heptonstall, Heurs, Hild, Hinderer,
  Hoak, Hofman, Holt, Holz, Hopkins, Horst, Hough, Houston, Howell, Hreibi, Hu,
  Huerta, Huet, Hughey, Husa, Huttner, Huynh-Dinh, Indik, Inta, Intini, Isa,
  Isac, Isi, Iyer, Izumi, Jacqmin, Jani, Jaranowski, Jawahar,
  Jim{\'{e}}nez-Forteza, Johnson, Johnson-McDaniel, Jones, Jones, Jonker, Ju,
  Junker, Kalaghatgi, Kalogera, Kamai, Kandhasamy, Kang, Kanner, Kapadia,
  Karki, Karvinen, Kasprzack, Kastaun, Katolik, Katsavounidis, Katzman, Kaufer,
  Kawabe, K{\'{e}}f{\'{e}}lian, Keitel, Kemball, Kennedy, Kent, Key, Khalili,
  Khan, Khan, Khan, Khazanov, Kijbunchoo, Kim, Kim, Kim, Kim, Kim, Kim,
  Kimbrell, King, King, Kinley-Hanlon, Kirchhoff, Kissel, Kleybolte, Klimenko,
  Knowles, Koch, Koehlenbeck, Koley, Kondrashov, Kontos, Korobko, Korth,
  Kowalska, Kozak, Krämer, Kringel, Krishnan, Kr{\'{o}}lak, Kuehn, Kumar,
  Kumar, Kumar, Kuo, Kutynia, Kwang, Lackey, Lai, Landry, Lang, Lange, Lantz,
  Lanza, Lartaux-Vollard, Lasky, Laxen, Lazzarini, Lazzaro, Leaci, Leavey, Lee,
  Lee, Lee, Lee, Lee, Lehmann, Lenon, Leonardi, Leroy, Letendre, Levin, Li,
  Linker, Littenberg, Liu, Lo, Lockerbie, London, Lord, Lorenzini, Loriette,
  Lormand, Losurdo, Lough, Lousto, Lovelace, Lück, Lumaca, Lundgren, Lynch,
  Ma, Macas, Macfoy, Machenschalk, MacInnis, Macleod, Hernandez,
  Maga{\~{n}}a-Sandoval, Zertuche, Magee, Majorana, Maksimovic, Man, Mandic,
  Mangano, Mansell, Manske, Mantovani, Marchesoni, Marion, M{\'{a}}rka,
  M{\'{a}}rka, Markakis, Markosyan, Markowitz, Maros, Marquina, Martelli,
  Martellini, Martin, Martin, Martynov, Mason, Massera, Masserot, Massinger,
  Masso-Reid, Mastrogiovanni, Matas, Matichard, Matone, Mavalvala, Mazumder,
  McCarthy, McClelland, McCormick, McCuller, McGuire, McIntyre, McIver,
  McManus, McNeill, McRae, McWilliams, Meacher, Meadors, Mehmet, Meidam,
  Mejuto-Villa, Melatos, Mendell, Mercer, Merilh, Merzougui, Meshkov,
  Messenger, Messick, Metzdorff, Meyers, Miao, Michel, Middleton, Mikhailov,
  Milano, Miller, Miller, Miller, Millhouse, Milovich-Goff, Minazzoli,
  Minenkov, Ming, Mishra, Mitra, Mitrofanov, Mitselmakher, Mittleman, Moffa,
  Moggi, Mogushi, Mohan, Mohapatra, Montani, Moore, Moraru, Moreno, Morriss,
  Mours, Mow-Lowry, Mueller, Muir, Mukherjee, Mukherjee, Mukherjee, Mukund,
  Mullavey, Munch, Mu{\~{n}}iz, Muratore, Murray, Napier, Nardecchia,
  Naticchioni, Nayak, Neilson, Nelemans, Nelson, Nery, Neunzert, Nevin,
  Newport, Newton, Ng, Nguyen, Nichols, Nielsen, Nissanke, Nitz, Noack, Nocera,
  Nolting, North, Nuttall, Oberling, O'Dea, Ogin, Oh, Oh, Ohme, Okada, Oliver,
  Oppermann, Oram, O'Reilly, Ormiston, Ortega, O'Shaughnessy, Ossokine,
  Ottaway, Overmier, Owen, Pace, Page, Page, Pai, Pai, Palamos, Palashov,
  Palomba, Pal-Singh, Pan, Pan, Pang, Pang, Pankow, Pannarale, Pant, Paoletti,
  Paoli, Papa, Parida, Parker, Pascucci, Pasqualetti, Passaquieti, Passuello,
  Patil, Patricelli, Pearlstone, Pedraza, Pedurand, Pekowsky, Pele, Penn,
  Perez, Perreca, Perri, Pfeiffer, Phelps, Piccinni, Pichot, Piergiovanni,
  Pierro, Pillant, Pinard, Pinto, Pirello, Pitkin, Poe, Poggiani, Popolizio,
  Porter, Post, Powell, Prasad, Pratt, Pratten, Predoi, Prestegard, Prijatelj,
  Principe, Privitera, Prodi, Prokhorov, Puncken, Punturo, Puppo, Pürrer, Qi,
  Quetschke, Quintero, Quitzow-James, Raab, Rabeling, Radkins, Raffai, Raja,
  Rajan, Rajbhandari, Rakhmanov, Ramirez, Ramos-Buades, Rapagnani, Raymond,
  Razzano, Read, Regimbau, Rei, Reid, Reitze, Ren, Reyes, Ricci, Ricker,
  Rieger, Riles, Rizzo, Robertson, Robie, Robinet, Rocchi, Rolland, Rollins,
  Roma, Romano, Romel, Romie, Rosi{\'{n}}ska, Ross, Rowan, Rüdiger, Ruggi,
  Rutins, Ryan, Sachdev, Sadecki, Sadeghian, Sakellariadou, Salconi, Saleem,
  Salemi, Samajdar, Sammut, Sampson, Sanchez, Sanchez, Sanchis-Gual, Sandberg,
  Sanders, Sassolas, Sathyaprakash, Saulson, Sauter, Savage, Sawadsky, Schale,
  Scheel, Scheuer, Schmidt, Schmidt, Schnabel, Schofield, Schönbeck,
  Schreiber, Schuette, Schulte, Schutz, Schwalbe, Scott, Scott, Seidel,
  Sellers, Sengupta, Sentenac, Sequino, Sergeev, Shaddock, Shaffer, Shah,
  Shahriar, Shaner, Shao, Shapiro, Shawhan, Sheperd, Shoemaker, Shoemaker,
  Siellez, Siemens, Sieniawska, Sigg, Silva, Singer, Singh, Singhal, Sintes,
  Slagmolen, Smith, Smith, Smith, Somala, Son, Sonnenberg, Sorazu, Sorrentino,
  Souradeep, Spencer, Srivastava, Staats, Staley, Steinke, Steinlechner,
  Steinlechner, Steinmeyer, Stevenson, Stone, Stops, Strain, Stratta, Strigin,
  Strunk, Sturani, Stuver, Summerscales, Sun, Sunil, Suresh, Sutton, Swinkels,
  Szczepa{\'{n}}czyk, Tacca, Tait, Talbot, Talukder, Tanner, T{\'{a}}pai,
  Taracchini, Tasson, Taylor, Taylor, Tewari, Theeg, Thies, Thomas, Thomas,
  Thomas, Thorne, Thorne, Thrane, Tiwari, Tiwari, Tokmakov, Toland, Tonelli,
  Tornasi, Torres-Forn{\'{e}}, Torrie, Töyrä, Travasso, Traylor, Trinastic,
  Tringali, Trozzo, Tsang, Tse, Tso, Tsukada, Tsuna, Tuyenbayev, Ueno, Ugolini,
  Unnikrishnan, Urban, Usman, Vahlbruch, Vajente, Valdes, van Bakel, van
  Beuzekom, van~den Brand, Broeck, Vander-Hyde, van~der Schaaf, van Heijningen,
  van Veggel, Vardaro, Varma, Vass, Vas{\'{u}}th, Vecchio, Vedovato, Veitch,
  Veitch, Venkateswara, Venugopalan, Verkindt, Vetrano, Vicer{\'{e}}, Viets,
  Vinciguerra, Vine, Vinet, Vitale, Vo, Vocca, Vorvick, Vyatchanin, Wade, Wade,
  Wade, Walet, Walker, Wallace, Walsh, Wang, Wang, Wang, Wang, Wang, Ward,
  Warner, Was, Watchi, Weaver, Wei, Weinert, Weinstein, Weiss, Wen, Wessel,
  We{\ss}els, Westerweck, Westphal, Wette, Whelan, Whitcomb, Whiting, Whittle,
  Wilken, Williams, Williams, Williamson, Willis, Willke, Wimmer, Winkler,
  Wipf, Wittel, Woan, Woehler, Wofford, Wong, Worden, Wright, Wu, Wysocki,
  Xiao, Yamamoto, Yancey, Yang, Yap, Yazback, Yu, Yu, Yvert, Zadro{\.{z}}ny,
  Zanolin, Zelenova, Zendri, Zevin, Zhang, Zhang, Zhang, Zhang, Zhao, Zhou,
  Zhou, Zhu, Zhu, Zimmerman, Zucker, Zweizig, Burns, Veres, Kocevski, Racusin,
  Goldstein, Connaughton, Briggs, Blackburn, Hamburg, Hui, von Kienlin,
  McEnery, Preece, Wilson-Hodge, Bissaldi, Cleveland, Gibby, Giles, Kippen,
  McBreen, Meegan, Paciesas, Poolakkil, Roberts, Stanbro, Savchenko, Ferrigno,
  Kuulkers, Bazzano, Bozzo, Brandt, Chenevez, Courvoisier, Diehl, Domingo,
  Hanlon, Jourdain, Laurent, Lebrun, Lutovinov, Mereghetti, Natalucci, Rodi,
  Roques, Sunyaev, Ubertini, , \& and}]{gw170817-speed}
Abbott, B.~P., Abbott, R., Abbott, T.~D., {et~al.} 2017{\natexlab{a}}, The
  Astrophysical Journal, 848, L13, \dodoi{10.3847/2041-8213/aa920c}

\bibitem[{Abbott {et~al.}(2017{\natexlab{b}})Abbott, Abbott, Abbott, Abernathy,
  Ackley, Adams, Addesso, Adhikari, Adya, Affeldt, Aggarwal, Aguiar, Ain,
  Ajith, Allen, Altin, Anderson, Anderson, Arai, Araya, Arceneaux, Areeda,
  Arun, Ashton, Ast, Aston, Aufmuth, Aulbert, Babak, Baker, Ballmer, Barayoga,
  Barclay, Barish, Barker, Barr, Barsotti, Bartlett, Bartos, Bassiri, Batch,
  Baune, Bell, Berger, Bergmann, Berry, Betzwieser, Bhagwat, Bhandare, Bilenko,
  Billingsley, Birch, Birney, Biscans, Bisht, Biwer, Blackburn, Blair, Blair,
  Blair, Bock, Bogan, Bohe, Bond, Bork, Bose, Brady, Braginsky, Brau,
  Brinkmann, Brockill, Broida, Brooks, Brown, Brown, Brown, Brunett, Buchanan,
  Buikema, Buonanno, Byer, Cabero, Cadonati, Cahillane, Bustillo, Callister,
  Camp, Cannon, Cao, Capano, Caride, Caudill, Cavagli{\`{a}}, Cepeda,
  Chamberlin, Chan, Chao, Charlton, Cheeseboro, Chen, Chen, Cheng, Cho, Cho,
  Chow, Christensen, Chu, Chung, Ciani, Clara, Clark, Collette, Cominsky,
  Constancio, Cook, Corbitt, Cornish, Corsi, Costa, Coughlin, Coughlin,
  Countryman, Couvares, Cowan, Coward, Cowart, Coyne, Coyne, Craig, Creighton,
  Cripe, Crowder, Cumming, Cunningham, Canton, Danilishin, Danzmann, Darman,
  Dasgupta, Costa, Dave, Davies, Daw, De, DeBra, Pozzo, Denker, Dent,
  Dergachev, DeRosa, DeSalvo, Devine, Dhurandhar, D{\'{\i}}az, Palma, Donovan,
  Dooley, Doravari, Douglas, Downes, Drago, Drever, Driggers, Dwyer, Edo,
  Edwards, Effler, Eggenstein, Ehrens, Eichholz, Eikenberry, Engels, Essick,
  Etzel, Evans, Evans, Everett, Factourovich, Fair, Fairhurst, Fan, Fang, Farr,
  Farr, Favata, Fays, Fehrmann, Fejer, Fenyvesi, Ferreira, Fisher, Fletcher,
  Frei, Freise, Frey, Fritschel, Frolov, Fulda, Fyffe, Gabbard, Gair, Gaonkar,
  Gaur, Gehrels, Geng, George, Gergely, Ghosh, Ghosh, Giaime, Giardina, Gill,
  Glaefke, Goetz, Goetz, Gondan, Gonz{\'{a}}lez, Gopakumar, Gordon, Gorodetsky,
  Gossan, Graef, Graff, Grant, Gras, Gray, Green, Grote, Grunewald, Guo, Gupta,
  Gupta, Gushwa, Gustafson, Gustafson, Hacker, Hall, Hall, Hammond, Haney,
  Hanke, Hanks, Hanna, Hannam, Hanson, Hardwick, Harry, Harry, Hart, Hartman,
  Haster, Haughian, Heintze, Hendry, Heng, Hennig, Henry, Heptonstall, Heurs,
  Hild, Hoak, Holt, Holz, Hopkins, Hough, Houston, Howell, Hu, Huang, Huerta,
  Hughey, Husa, Huttner, Huynh-Dinh, Indik, Ingram, Inta, Isa, Isi, Isogai,
  Iyer, Izumi, Jang, Jani, Jawahar, Jian, Jim{\'{e}}nez-Forteza, Johnson,
  Jones, Jones, Ju, Haris, Kalaghatgi, Kalogera, Kandhasamy, Kang, Kanner,
  Kapadia, Karki, Karvinen, Kasprzack, Katsavounidis, Katzman, Kaufer, Kaur,
  Kawabe, Kehl, Keitel, Kelley, Kells, Kennedy, Key, Khalili, Khan, Khan,
  Khazanov, Kijbunchoo, Kim, Kim, Kim, Kim, Kim, Kim, Kim, Kimbrell, King,
  King, Kissel, Klein, Kleybolte, Klimenko, Koehlenbeck, Kondrashov, Kontos,
  Korobko, Korth, Kozak, Kringel, Krueger, Kuehn, Kumar, Kumar, Kuo, Lackey,
  Landry, Lange, Lantz, Lasky, Laxen, Lazzarini, Leavey, Lebigot, Lee, Lee,
  Lee, Lee, Lenon, Leong, Levin, Lewis, Li, Libson, Littenberg, Lockerbie,
  Lombardi, London, Lord, Lormand, Lough, Lück, Lundgren, Lynch, Ma,
  Machenschalk, MacInnis, Macleod, Maga{\~{n}}a-Sandoval, Zertuche, Magee,
  Mandic, Mangano, Mansell, Manske, M{\'{a}}rka, M{\'{a}}rka, Markosyan, Maros,
  Martin, Martynov, Mason, Massinger, Masso-Reid, Matichard, Matone, Mavalvala,
  Mazumder, McCarthy, McClelland, McCormick, McGuire, McIntyre, McIver,
  McManus, McRae, McWilliams, Meacher, Meadors, Melatos, Mendell, Mercer,
  Merilh, Meshkov, Messenger, Messick, Meyers, Miao, Middleton, Mikhailov,
  Miller, Miller, Miller, Miller, Millhouse, Ming, Mirshekari, Mishra, Mitra,
  Mitrofanov, Mitselmakher, Mittleman, Mohapatra, Moore, Moore, Moraru, Moreno,
  Morriss, Mossavi, Mow-Lowry, Mueller, Muir, Mukherjee, Mukherjee, Mukherjee,
  Mukund, Mullavey, Munch, Murphy, Murray, Mytidis, Nayak, Nedkova, Nelson,
  Neunzert, Newton, Nguyen, Nielsen, Nitz, Nolting, Normandin, Nuttall,
  Oberling, Ochsner, O'Dell, Oelker, Ogin, Oh, Oh, Ohme, Oliver, Oppermann,
  Oram, O'Reilly, O'Shaughnessy, Ottaway, Overmier, Owen, Pai, Pai, Palamos,
  Palashov, Pal-Singh, Pan, Pankow, Pannarale, Pant, Papa, Paris, Parker,
  Pascucci, Patrick, Pearlstone, Pedraza, Pekowsky, Pele, Penn, Perreca, Perri,
  Phelps, Pierro, Pinto, Pitkin, Poe, Post, Powell, Prasad, Predoi, Prestegard,
  Price, Prijatelj, Principe, Privitera, Prokhorov, Puncken, Pürrer, Qi, Qin,
  Qiu, Quetschke, Quintero, Quitzow-James, Raab, Rabeling, Radkins, Raffai,
  Raja, Rajan, Rakhmanov, Raymond, Read, Reed, Reid, Reitze, Rew, Reyes, Riles,
  Rizzo, Robertson, Robie, Rollins, Roma, Romanov, Romie, Rowan, Rüdiger,
  Ryan, Sachdev, Sadecki, Sadeghian, Sakellariadou, Saleem, Salemi, Samajdar,
  Sammut, Sanchez, Sandberg, Sandeen, Sanders, Sathyaprakash, Saulson, Sauter,
  Savage, Sawadsky, Schale, Schilling, Schmidt, Schmidt, Schnabel, Schofield,
  Schönbeck, Schreiber, Schuette, Schutz, Scott, Scott, Sellers, Sengupta,
  Sergeev, Shaddock, Shaffer, Shahriar, Shaltev, Shapiro, Shawhan, Sheperd,
  Shoemaker, Shoemaker, Siellez, Siemens, Sigg, Silva, Singer, Singer, Singh,
  Singh, Sintes, Slagmolen, Smith, Smith, Smith, Son, Sorazu, Souradeep,
  Srivastava, Staley, Steinke, Steinlechner, Steinlechner, Steinmeyer,
  Stephens, Stone, Strain, Strauss, Strigin, Sturani, Stuver, Summerscales,
  Sun, Sunil, Sutton, Szczepa{\'{n}}czyk, Talukder, Tanner, T{\'{a}}pai,
  Tarabrin, Taracchini, Taylor, Theeg, Thirugnanasambandam, Thomas, Thomas,
  Thomas, Thorne, Thrane, Tiwari, Tokmakov, Toland, Tomlinson, Tornasi, Torres,
  Torrie, Töyrä, Traylor, Trifir{\`{o}}, Tse, Tuyenbayev, Ugolini,
  Unnikrishnan, Urban, Usman, Vahlbruch, Vajente, Valdes, Vander-Hyde, van
  Veggel, Vass, Vaulin, Vecchio, Veitch, Veitch, Venkateswara, Vinciguerra,
  Vine, Vitale, Vo, Vorvick, Voss, Vousden, Vyatchanin, Wade, Wade, Wade,
  Walker, Wallace, Walsh, Wang, Wang, Wang, Wang, Ward, Warner, Weaver,
  Weinert, Weinstein, Weiss, Wen, We{\ss}els, Westphal, Wette, Whelan, Whiting,
  Williams, Williamson, Willis, Willke, Wimmer, Winkler, Wipf, Wittel, Woan,
  Woehler, Worden, Wright, Wu, Wu, Yablon, Yam, Yamamoto, Yancey, Yu, Zanolin,
  Zevin, Zhang, Zhang, Zhang, Zhao, Zhou, Zhou, Zhu, Zucker, Zuraw, Zweizig, \&
  and}]{ce}
---. 2017{\natexlab{b}}, Classical and Quantum Gravity, 34, 044001,
  \dodoi{10.1088/1361-6382/aa51f4}

\bibitem[{Abbott {et~al.}(2018)}]{livingreviewligo}
Abbott, B.~P., {et~al.} 2018, Living Rev. Rel., 21, 3,
  \dodoi{10.1007/s41114-018-0012-9, 10.1007/lrr-2016-1}

\bibitem[{Abbott {et~al.}(2019{\natexlab{a}})}]{GWTC1-testingGR}
---. 2019{\natexlab{a}}, Phys. Rev. D, 100, 104036,
  \dodoi{10.1103/PhysRevD.100.104036}

\bibitem[{Abbott {et~al.}(2019{\natexlab{b}})}]{gwtc}
---. 2019{\natexlab{b}}, Phys. Rev. X, 9, 031040,
  \dodoi{10.1103/PhysRevX.9.031040}

\bibitem[{Abbott {et~al.}(2019{\natexlab{c}})}]{GWTC1-rate}
---. 2019{\natexlab{c}}, Astrophys. J. Lett., 882, L24,
  \dodoi{10.3847/2041-8213/ab3800}

\bibitem[{Abbott {et~al.}(2020)}]{GW190425}
---. 2020, Astrophys. J. Lett., 892, L3, \dodoi{10.3847/2041-8213/ab75f5}

\bibitem[{Aghanim {et~al.}(2020)}]{Planck2018}
Aghanim, N., {et~al.} 2020, Astron. Astrophys., 641, A6,
  \dodoi{10.1051/0004-6361/201833910}

\bibitem[{Alexander \& Yunes(2009)}]{cs}
Alexander, S., \& Yunes, N. 2009, Physics Reports, 480, 1 ,
  \dodoi{https://doi.org/10.1016/j.physrep.2009.07.002}

\bibitem[{Alexander \& Yunes(2018)}]{yunes2018}
Alexander, S.~H., \& Yunes, N. 2018, Phys. Rev. D, 97, 064033,
  \dodoi{10.1103/PhysRevD.97.064033}

\bibitem[{Ali-Ha\"{\i}moud(2011)}]{pulsar2}
Ali-Ha\"{\i}moud, Y. 2011, Phys. Rev. D, 83, 124050,
  \dodoi{10.1103/PhysRevD.83.124050}

\bibitem[{Arai \& Nishizawa(2018)}]{Arai:2017hxj}
Arai, S., \& Nishizawa, A. 2018, Phys. Rev. D, 97, 104038,
  \dodoi{10.1103/PhysRevD.97.104038}

\bibitem[{Berti {et~al.}(2015)Berti, Barausse, Cardoso, Gualtieri, Pani,
  Sperhake, Stein, Wex, Yagi, Baker, Burgess, Coelho, Doneva, Felice, Ferreira,
  Freire, Healy, Herdeiro, Horbatsch, Kleihaus, Klein, Kokkotas, Kunz, Laguna,
  Lang, Li, Littenberg, Matas, Mirshekari, Okawa, Radu, O'Shaughnessy,
  Sathyaprakash, Broeck, Winther, Witek, Aghili, Alsing, Bolen, Bombelli,
  Caudill, Chen, Degollado, Fujita, Gao, Gerosa, Kamali, Silva, Rosa,
  Sadeghian, Sampaio, Sotani, \& Zilhao}]{a2}
Berti, E., Barausse, E., Cardoso, V., {et~al.} 2015, Classical and Quantum
  Gravity, 32, 243001, \dodoi{10.1088/0264-9381/32/24/243001}

\bibitem[{Biwer {et~al.}(2019)Biwer, Capano, De, Cabero, Brown, Nitz, \&
  Raymond}]{pycbcinference}
Biwer, C.~M., Capano, C.~D., De, S., {et~al.} 2019, Publications of the
  Astronomical Society of the Pacific, 131, 024503,
  \dodoi{10.1088/1538-3873/aaef0b}

\bibitem[{Conroy \& Koivisto(2019)}]{tele}
Conroy, A., \& Koivisto, T. 2019, Journal of Cosmology and Astroparticle
  Physics, 2019, 016, \dodoi{10.1088/1475-7516/2019/12/016}

\bibitem[{Creminelli {et~al.}(2014)Creminelli, Gleyzes, Nore\~na, \&
  Vernizzi}]{eft}
Creminelli, P., Gleyzes, J., Nore\~na, J., \& Vernizzi, F. 2014, Phys. Rev.
  Lett., 113, 231301, \dodoi{10.1103/PhysRevLett.113.231301}

\bibitem[{Crisostomi {et~al.}(2018)Crisostomi, Noui, Charmousis, \&
  Langlois}]{ghost}
Crisostomi, M., Noui, K., Charmousis, C., \& Langlois, D. 2018, Phys. Rev. D,
  97, 044034, \dodoi{10.1103/PhysRevD.97.044034}

\bibitem[{Dietrich {et~al.}(2017)Dietrich, Bernuzzi, \& Tichy}]{nrtidal}
Dietrich, T., Bernuzzi, S., \& Tichy, W. 2017, Phys. Rev. D, 96, 121501,
  \dodoi{10.1103/PhysRevD.96.121501}

\bibitem[{Hannam {et~al.}(2014)Hannam, Schmidt, Boh\'e, Haegel, Husa, Ohme,
  Pratten, \& P\"urrer}]{pv22}
Hannam, M., Schmidt, P., Boh\'e, A., {et~al.} 2014, Phys. Rev. Lett., 113,
  151101, \dodoi{10.1103/PhysRevLett.113.151101}

\bibitem[{Ho\ifmmode~\check{r}\else \v{r}\fi{}ava(2009)}]{HL0}
Ho\ifmmode~\check{r}\else \v{r}\fi{}ava, P. 2009, Phys. Rev. D, 79, 084008,
  \dodoi{10.1103/PhysRevD.79.084008}

\bibitem[{Horndeski(1974)}]{Horndeski:1974wa}
Horndeski, G.~W. 1974, Int. J. Theor. Phys., 10, 363,
  \dodoi{10.1007/BF01807638}

\bibitem[{Kostelecky(2004)}]{Kostelecky:2003fs}
Kostelecky, V. 2004, Phys. Rev. D, 69, 105009,
  \dodoi{10.1103/PhysRevD.69.105009}

\bibitem[{Kostelecky \& Russell(2011)}]{Kostelecky:2008ts}
Kostelecky, V., \& Russell, N. 2011, Rev. Mod. Phys., 83, 11,
  \dodoi{10.1103/RevModPhys.83.11}

\bibitem[{Kosteleck{\'y} \& Mewes(2016)}]{mewes}
Kosteleck{\'y}, V.~A., \& Mewes, M. 2016, Physics Letters B, 757, 510 ,
  \dodoi{https://doi.org/10.1016/j.physletb.2016.04.040}

\bibitem[{Kosteleck\'y \& Mewes(2018)}]{Kostelecky:2017zob}
Kosteleck\'y, V.~A., \& Mewes, M. 2018, Phys. Lett. B, 779, 136,
  \dodoi{10.1016/j.physletb.2018.01.082}

\bibitem[{Lee \& Yang(1956)}]{lee-yang}
Lee, T.~D., \& Yang, C.~N. 1956, Phys. Rev., 104, 254,
  \dodoi{10.1103/PhysRev.104.254}

\bibitem[{{LIGO Scientific Collaboration}(2018)}]{lalsuite}
{LIGO Scientific Collaboration}. 2018, {LIGO} {A}lgorithm {L}ibrary -
  {LALS}uite, free software (GPL), \dodoi{10.7935/GT1W-FZ16}

\bibitem[{Miller \& Yunes(2019)}]{miller_2019}
Miller, M.~C., \& Yunes, N. 2019, Nature, 568, 469,
  \dodoi{10.1038/s41586-019-1129-z}

\bibitem[{Nair {et~al.}(2019)Nair, Perkins, Silva, \& Yunes}]{Nair:2019iur}
Nair, R., Perkins, S., Silva, H.~O., \& Yunes, N. 2019, Phys. Rev. Lett., 123,
  191101, \dodoi{10.1103/PhysRevLett.123.191101}

\bibitem[{Nishizawa(2018)}]{Nishizawa:2017nef}
Nishizawa, A. 2018, Phys. Rev. D, 97, 104037,
  \dodoi{10.1103/PhysRevD.97.104037}

\bibitem[{Nishizawa \& Arai(2019)}]{Nishizawa:2019rra}
Nishizawa, A., \& Arai, S. 2019, Phys. Rev. D, 99, 104038,
  \dodoi{10.1103/PhysRevD.99.104038}

\bibitem[{Nishizawa \& Kobayashi(2018)}]{1809}
Nishizawa, A., \& Kobayashi, T. 2018, Phys. Rev. D, 98, 124018,
  \dodoi{10.1103/PhysRevD.98.124018}

\bibitem[{Punturo {et~al.}(2010)Punturo, Abernathy, Acernese, Allen, Andersson,
  Arun, Barone, Barr, Barsuglia, Beker, Beveridge, Birindelli, Bose, Bosi,
  Braccini, Bradaschia, Bulik, Calloni, Cella, Mottin, Chelkowski, Chincarini,
  Clark, Coccia, Colacino, Colas, Cumming, Cunningham, Cuoco, Danilishin,
  Danzmann, Luca, Salvo, Dent, Rosa, Fiore, Virgilio, Doets, Fafone, Falferi,
  Flaminio, Franc, Frasconi, Freise, Fulda, Gair, Gemme, Gennai, Giazotto,
  Glampedakis, Granata, Grote, Guidi, Hammond, Hannam, Harms, Heinert, Hendry,
  Heng, Hennes, Hild, Hough, Husa, Huttner, Jones, Khalili, Kokeyama, Kokkotas,
  Krishnan, Lorenzini, Lück, Majorana, Mandel, Mandic, Martin, Michel,
  Minenkov, Morgado, Mosca, Mours, Müller{\textendash}Ebhardt, Murray,
  Nawrodt, Nelson, Oshaughnessy, Ott, Palomba, Paoli, Parguez, Pasqualetti,
  Passaquieti, Passuello, Pinard, Poggiani, Popolizio, Prato, Puppo, Rabeling,
  Rapagnani, Read, Regimbau, Rehbein, Reid, Rezzolla, Ricci, Richard, Rocchi,
  Rowan, Rüdiger, Sassolas, Sathyaprakash, Schnabel, Schwarz, Seidel, Sintes,
  Somiya, Speirits, Strain, Strigin, Sutton, Tarabrin, Thüring, van~den Brand,
  van Leewen, van Veggel, van~den Broeck, Vecchio, Veitch, Vetrano, Vicere,
  Vyatchanin, Willke, Woan, Wolfango, \& Yamamoto}]{et}
Punturo, M., Abernathy, M., Acernese, F., {et~al.} 2010, Classical and Quantum
  Gravity, 27, 194002, \dodoi{10.1088/0264-9381/27/19/194002}

\bibitem[{Reitze {et~al.}(2019)}]{cewhitepaper}
Reitze, D., {et~al.} 2019, Bull. Am. Astron. Soc., 51, 035.
\newblock \doarXiv{1907.04833}

\bibitem[{Schmidt {et~al.}(2015)Schmidt, Ohme, \& Hannam}]{pv2}
Schmidt, P., Ohme, F., \& Hannam, M. 2015, Phys. Rev. D, 91, 024043,
  \dodoi{10.1103/PhysRevD.91.024043}

\bibitem[{Smith {et~al.}(2008)Smith, Erickcek, Caldwell, \&
  Kamionkowski}]{solar}
Smith, T.~L., Erickcek, A.~L., Caldwell, R.~R., \& Kamionkowski, M. 2008, Phys.
  Rev. D, 77, 024015, \dodoi{10.1103/PhysRevD.77.024015}

\bibitem[{Speagle(2020)}]{dynesty}
Speagle, J.~S. 2020, Monthly Notices of the Royal Astronomical Society,
  \dodoi{10.1093/mnras/staa278}

\bibitem[{Vallisneri {et~al.}(2015)Vallisneri, Kanner, Williams, Weinstein, \&
  Stephens}]{gwosc}
Vallisneri, M., Kanner, J., Williams, R., Weinstein, A., \& Stephens, B. 2015,
  J. Phys. Conf. Ser., 610, 012021, \dodoi{10.1088/1742-6596/610/1/012021}

\bibitem[{Wang {et~al.}(2013)Wang, Wu, Zhao, \& Zhu}]{HL}
Wang, A., Wu, Q., Zhao, W., \& Zhu, T. 2013, Phys. Rev. D, 87, 103512,
  \dodoi{10.1103/PhysRevD.87.103512}

\bibitem[{Wu {et~al.}(1957)Wu, Ambler, Hayward, Hoppes, \& Hudson}]{cswuparity}
Wu, C.~S., Ambler, E., Hayward, R.~W., Hoppes, D.~D., \& Hudson, R.~P. 1957,
  Phys. Rev., 105, 1413, \dodoi{10.1103/PhysRev.105.1413}

\bibitem[{Yagi \& Yang(2018)}]{cs4}
Yagi, K., \& Yang, H. 2018, Phys. Rev. D, 97, 104018,
  \dodoi{10.1103/PhysRevD.97.104018}

\bibitem[{Yunes \& Siemens(2013)}]{a1}
Yunes, N., \& Siemens, X. 2013, Living Reviews in Relativity, 16,
  \dodoi{10.12942/lrr-2013-9}

\bibitem[{Yunes \& Spergel(2009)}]{pulsar}
Yunes, N., \& Spergel, D.~N. 2009, Phys. Rev. D, 80, 042004,
  \dodoi{10.1103/PhysRevD.80.042004}

\bibitem[{Zhao {et~al.}(2020{\natexlab{a}})Zhao, Liu, Wen, Zhu, Wang, Hu, \&
  Zhou}]{Zhao:2019}
Zhao, W., Liu, T., Wen, L., {et~al.} 2020{\natexlab{a}}, Eur. Phys. J. C, 80,
  630, \dodoi{10.1140/epjc/s10052-020-8211-4}

\bibitem[{Zhao {et~al.}(2020{\natexlab{b}})Zhao, Zhu, Qiao, \& Wang}]{zhao2020}
Zhao, W., Zhu, T., Qiao, J., \& Wang, A. 2020{\natexlab{b}}, Phys. Rev. D, 101,
  024002, \dodoi{10.1103/PhysRevD.101.024002}

\end{thebibliography}

\end{document}